\documentclass[pra,twocolumn,superscriptaddress,showpacs,floatfix,longbibliography]{revtex4-2}
\usepackage{mathrsfs,braket}
\usepackage{amssymb, amsbsy, amsmath, latexsym, dsfont, array, layout,
graphicx,mathrsfs,color,ulem,bm}
\usepackage[colorlinks=true,citecolor=blue,urlcolor=blue]{hyperref}

\begin{document}

\title{Dissipative two-dimensional Raman lattice}

\author{Haowei Li}
\affiliation{CAS Key Laboratory of Quantum Information, University of Science and Technology of China, Hefei 230026, China}
\author{Wei Yi}
\email{wyiz@ustc.edu.cn}
\affiliation{CAS Key Laboratory of Quantum Information, University of Science and Technology of China, Hefei 230026, China}
\affiliation{CAS Center For Excellence in Quantum Information and Quantum Physics, Hefei 230026, China}

\begin{abstract}
We show that a dissipative two-dimensional Raman lattice can be engineered in a two-component ultracold atomic gas, where the interplay of the two-dimensional spin-orbit coupling and light-induced atom loss gives rise to a density flow diagonal to the underlying square lattice. The flow is driven by the non-Hermitian corner skin effect, under which eigenstates localize toward one corner of the system.
We illustrate that the topological edge states of the system can only be accounted for by the non-Bloch band theory where the deformation of the bulk eigenstates are explicitly considered.
The directional flow can be detected through the dynamic evolution of an initially localized condensate in the lattice, or by introducing an immobile impurity species that interact spin-selectively with a condensate in the ground state of the Raman lattice.
\end{abstract}

\maketitle

\section{Introduction}

The recent experimental implementation of Raman lattices in cold atoms enables the simulation of topological matter such as topological insulators~\cite{xjliu1d,xjliu2d, shuai2d,topoins2, ti1} or Weyl semimetals~\cite{shuai3d,weyl2}.
While the excellent tunability of Raman lattices provides access to a plethora of dynamic topological phenomena either in quench processes or periodically-driven Floquet settings~\cite{dynRa1,dynRa2,dynRa3}, a largely unexplored possibility is the further introduction of dissipation. A dissipative Raman lattice should exhibit even more exotic properties in the highly non-trivial context of quantum open systems~\cite{dissRa}.
This is expected, since dissipation has been shown to stabilize interesting many-body phases ~\cite{zoller08,zoller12,blatt13} or phase transitions~\cite{disspt1,disspt2,disspt3,disspt4,disspt5} under the framework of quantum master equation. For the past several years, however, quantum open systems are often studied from the perspective of non-Hermitian physics: while the evolution of the density matrix in the quantum master equation is intrinsically non-Hermitian, state evolution under the condition of post selection (the so-called conditional dynamics) follows a non-Hermitian effective Hamiltonian~\cite{Non1,Uedareview,molmer,michael,weimer}. It is within this latter framework that non-Hermitian models with the parity-time symmetry~\cite{PT1,photonics2} or exotic non-Hermitian topology~\cite{nhtopot1,nhtopot2,nhtopot3,nhtopoe1,nhtopoe15,nhtopoe16} have been realized in cold atomic gases undergoing light-induced loss~\cite{luole,yanbo1,yanbo2}.

One of the most intriguing and fast-developing frontiers of non-Hermitian physics is the study of non-Hermitian skin effect (NHSE)~\cite{wangz1d,wangz2d,murakami,nhse1,nhse2,nhse3,nhse4,nhse5,nhse6,nhsedy1,nhsedy2,nhsedy3}. Defined as the exponential localization of eigenstates toward boundaries, the NHSE originates from a directional flow in the bulk that is closely related to the non-Hermitian model's spectral topology in the complex plane~\cite{nhse3,nhse4}. In a non-Hermitian lattice model with translational symmetry, since all eigenstates are deformed from Bloch waves, topological edge states can only be accounted for by the non-Bloch band theory, where the concept of the generalized Brillouin zone is introduced~\cite{wangz1d,wangz2d,murakami}. So far, dynamic signatures of the NHSE and its exotic implications for the band topology have been observed in quantum systems such as photons~\cite{nhtopoe2,scienceskin} and cold atoms~\cite{yanbo2}. These experiments, however, are confined to one dimension---discrete spatial modes for photons and distinct momentum states for atoms.
While higher-order NHSEs in higher spatial dimensions have been observed in acoutics~\cite{2dnhseacoustics} or topoelectrical circuits~\cite{2dnhsecircuits}, it is desirable to engineer non-Hermitian topological models in higher-dimensional quantum systems, where the NHSE has richer manifestations and impacts.

In this work, we propose a dissipative two-dimensional (2D) Raman lattice that is readily accessible in current experiments with ultracold atomic gases. An important ingredient of the model is the Raman-induced 2D spin-orbit coupling, under which the band topology of the lattice can be mapped to that of an anomalous quantum Hall insulator. We show that, by further imposing an on-site atom loss, the dissipative lattice acquires the non-Hermitian corner skin effect, where all eigenstates are exponentially localized toward a corner of the system under the open boundary condition. As a direct consequence, we confirm that topological invariants responsible for the topological edge states should be calculated over the generalized Brillouin zone, following the non-Bloch band theory.

Dynamically, the NHSE is reflected in a directional bulk density flow which vanishes in the absence of dissipation. The persist current forms the basis for previous experimental detection of the NHSE~\cite{nhtopoe2,yanbo2}, as well as its interplay with disorder~\cite{qw1,qw2}.
Here we propose two schemes to detect the bulk flow.
For concreteness, we consider a two-component Bose-Einstein condensate in the Raman lattice.
In the first scheme, the condensate is initialized in a quasi-local Gaussian wave packet over several sites deep in the bulk. The directional flow is visible in the subsequent time-evolved density distribution of the condensate.
In the second scheme, we consider a condensate initialized in the ground state of the Raman lattice. While the spatial homogeneity of the initial density distribution makes the directional flow invisible in the density evolution, we further introduce a third minority species (dubbed impurity) interacting spin-selectively with the condensate.
Through interactions, the impurity introduces spatially inhomogeneous density excitations in the background condensate, and the directional flow becomes visible from the dynamics of these excitations. We confirm the picture above through the coupled mean-field calculations, and demonstrate how the spin-selective interaction between the impurity and the condensate impacts the directional bulk flow.

The paper is organized as follows. In Sec.~II, we present the setup of the dissipative Raman lattice and derive the tight-binding Hamiltonian. We confirm the non-Hermitian corner skin effect and the non-Bloch topology of the system in Sec.~III. In Sec.~IV, we discuss in detail the two detection schemes.
We summarize in Sec.~V.

\begin{figure}[tbp]
	\includegraphics[width=8cm]{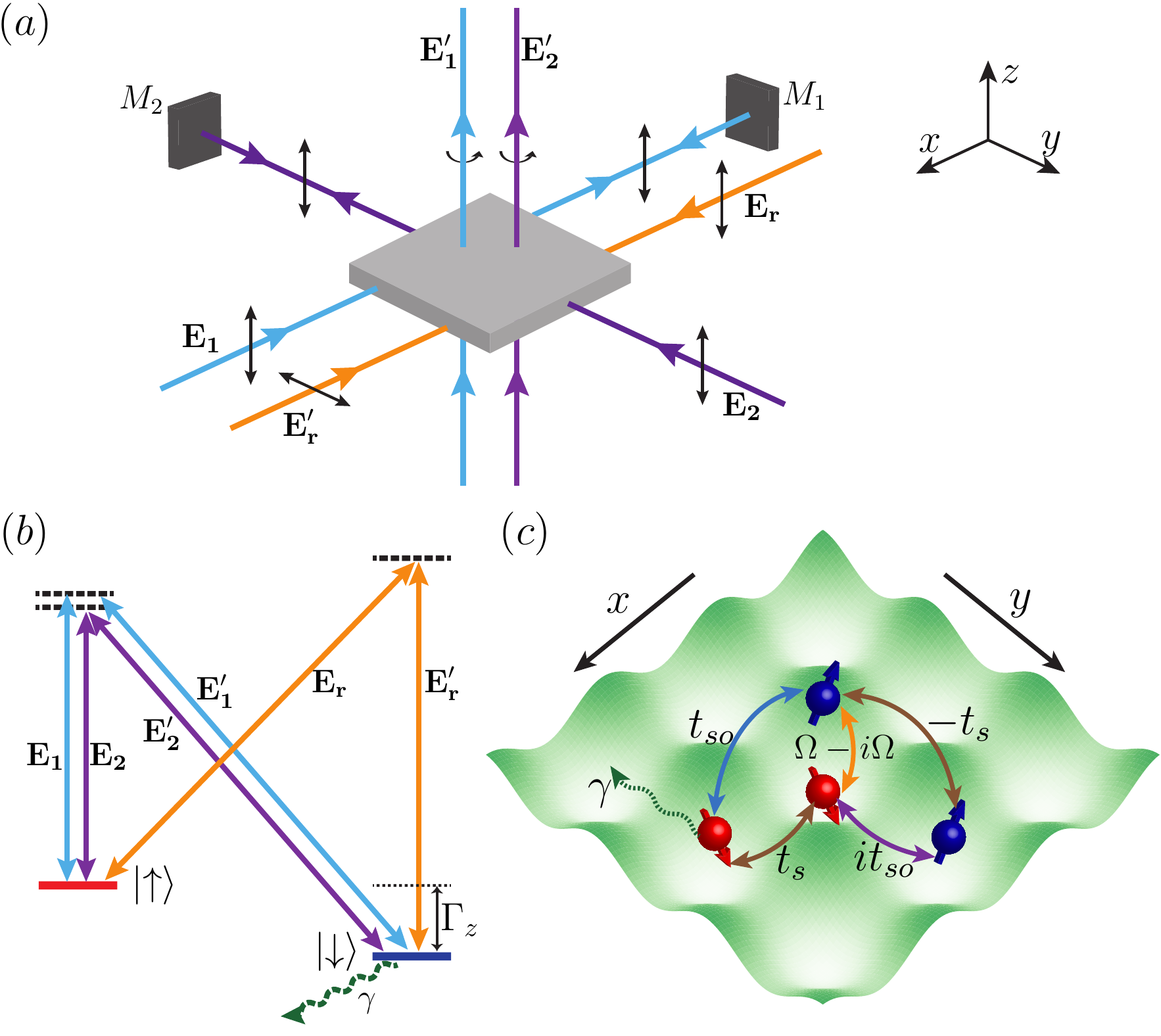}
	\caption{(a) Laser configuration for generating the Raman lattice. The black arrows indicate the laser polarizations. The standing-wave lasers $\mathbf{E_{1,2}}$ form a 2D optical lattice, and generate Raman couplings along the two spatial directions, together with the plane-wave lasers $\mathbf{E}^{\prime}_{1,2}$. The plane-wave lasers $\mathbf{E_{r}},\mathbf{E_{r}^\prime}$ induce on-site Raman couplings. (b) Level scheme: three sets of Raman lasers couple the states $\left|\uparrow\right\rangle$ and $\left|\downarrow\right\rangle$. (c) Schematic illustration of the dissipative 2D Raman lattice. See main text for the definition of the variables $\gamma$, $t_{so}$, $t_s$ and $\Omega$.}
	\label{fig:fig1}
\end{figure}

\begin{figure*}[tbp]
	\includegraphics[width=15cm]{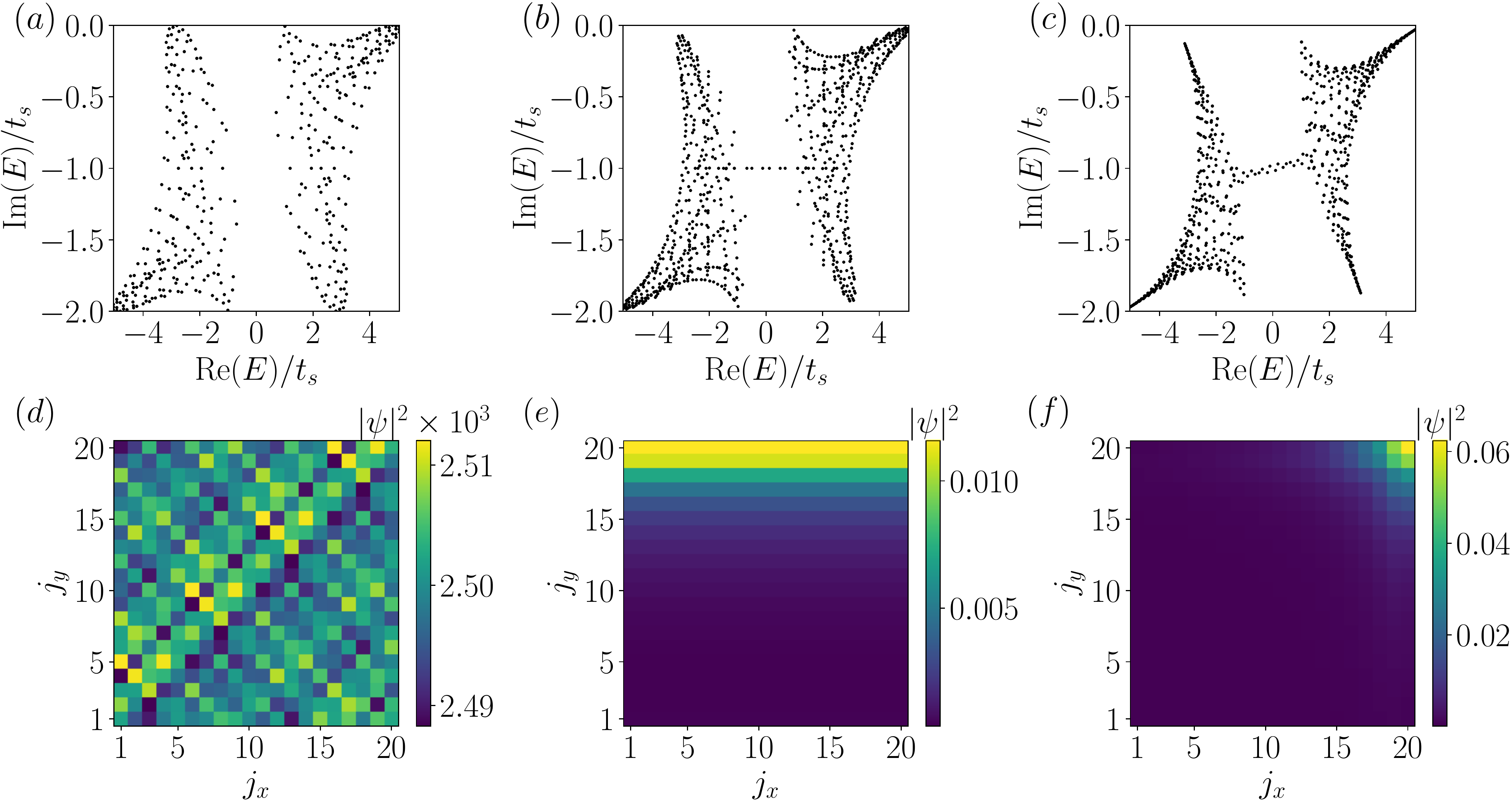}
\caption{(a)(b)(c) Eigenspectra of Hamiltonian (\ref{eq:H}) on the complex plane. (d)(e)(f) Average spatial distribution of all eigenstates of Hamiltonian (\ref{eq:H}). In (a)(d), the lattice is under the periodic boundary condition along both directions. In (b)(e), the lattice is under the periodic boundary condition along the $x$ direction, but open boundary condition along the $y$ direction. In (c)(f), the lattice is under the open boundary condition along both directions. For all figures, we take $t_{so}/t_s=1$, $\Gamma_{z}/t_s=1$, $\Omega/t_s=0.5$,  $\gamma/t_s=2$, and a lattice with $20\times 20$ sites. }
	\label{fig:fig2}
\end{figure*}

\section{Model}
As illustrated in Fig.~\ref{fig:fig1}, we consider a two-component ($\left|\uparrow\right\rangle$ and $\left|\downarrow\right\rangle$) atomic gas in a 2D optical lattice, described by the following Hamiltonian~\cite{xjliu2d,shuai2d}
\begin{equation}		
	\begin{aligned}
H(x,y)=&\frac{\mathbf{p}^2}{2m}+V(\mathbf{r})+\Gamma_z(\left|\uparrow\right\rangle\left\langle\uparrow\right|-\left|\downarrow\right\rangle\left\langle\downarrow\right|)-\mathrm{i}\gamma\left|\downarrow\right\rangle\left\langle\downarrow\right|\\&+\left\{\left[M_x(x)+\mathrm{i}M_y(y)+M_r(x)\right]\left|\uparrow\right\rangle\left\langle\downarrow\right|+\text{H.c.}\right\},
	\end{aligned}
	\label{eq:Hreal}
\end{equation}
where $\Gamma_{z}$ is the effective Zeeman field, $\gamma$ is the laser-induced loss rate of state $\left|\downarrow\right\rangle$.
{ Similar to the practice in Refs.~\cite{luole,yanbo2,joep}, the non-Hermitian term is introduced by coupling atoms in $\left|\downarrow\right\rangle$ to an electronically excited state which undergoes rapid spontaneous decay into the environment. The atoms that remain (within the lattice potential and in states $\left|\downarrow\right\rangle$ and $\left|\uparrow\right\rangle$) necessarily do not undergo the laser-induced decay process, and their dynamics is driven by the non-Hermitian effective Hamiltonian in (\ref{eq:Hreal}). For detailed derivations starting from the Lindblad master equation, see Refs.~\cite{nhsedy2,cuisoc}.}

The Raman lattice potential $V(\mathbf{r})=-V_0\left[\cos^2(k_0x)+\cos^2(k_0y)\right]$, and the Raman coupling terms $M_x(x)=M_0\sin(k_0x)$, $M_y(y)=M_0\sin(k_0y)$, and $M_r(x)=M_re^{\mathrm{i}(k_0x-\pi/4)}$.
Here the Raman lattice potential $V(\mathbf{r})$ is generated by the standing-wave lasers propagating along the $x$ and $y$ directions, with the electric fields $\mathbf{E_1}=\mathbf{e_z}E_{1}\sin(k_0 x)$ and $\mathbf{E_2}=\mathbf{e_z}E_{2}\sin(k_0 y)$, respectively. To introduce Raman couplings along the two spatial directions, we consider two plane-wave lasers (given by electric fields $\mathbf{E_1^\prime}$ and $\mathbf{E_2^\prime}$) with $\sigma^+$ polarization, propagating along the $z$ axis. The laser pairs $(\mathbf{E_1},\mathbf{E_1^\prime})$ and $(\mathbf{E_2},\mathbf{E_2^\prime})$ separately induce the terms $M_x(x)$ and $M_y(y)$ in Eq.~(\ref{eq:Hreal}). We further introduce the Raman coupling term $M_r(x)$ term in Eq.~(\ref{eq:Hreal}), by the plane-wave lasers $\mathbf{E_r}$ and $\mathbf{E_r^\prime}$ propagating along the $x$ axis, with electric fields $\mathbf{E_r}=\mathbf{e_y}E_{r}e^{\mathrm{i}k_0 x/2-\pi/4}$ and $\mathbf{E_r^\prime}=\mathbf{e_z}E_{r}^\prime e^{-\mathrm{i}k_0 x/2}$, respectively.

Under the single-band approximation, the tight-binding model corresponding to Eq.~(\ref{eq:Hreal}) can be derived with the lowest-band Wannier functions $\phi_\sigma^{j_x,j_y}(x,y)$, where $\sigma=\uparrow,\downarrow$. Here $j_x$ and $j_y$ are the  lattice-site labels in the $x$ and $y$ directions, both with a lattice spacing $a=\pi/k_0$. We only consider the on-site and nearest-neighbour hopping terms.
Following a gauge transformation ${c}_{\vec{j}\downarrow}\rightarrow e^{\mathrm{i}(j_x+j_y)}{c}_{\vec{j}\downarrow}$,
the tight-binding Hamiltonian is given by

\begin{equation}		
	\begin{aligned}
		H=&-t_{s} \sum_{<\vec{i}, \vec{j}>}\left({c}_{\vec{i}\uparrow}^{\dagger} {c}_{\vec{j}\uparrow}-{c}_{\vec{i} \downarrow}^{\dagger} {c}_{\vec{j} \downarrow}\right)+\sum_{\vec{i}} \Gamma_{z}\left({n}_{\vec{i} \uparrow}-{n}_{\vec{i} \downarrow}\right)\\
		&+\left[\sum_{\vec{i}} t_{\mathrm{so}}\left({c}_{\vec{i} \uparrow}^{\dagger} {c}_{\vec{i}+\vec{e}_x \downarrow}-{c}_{\vec{i} \uparrow}^{\dagger} {c}_{\vec{i}-\vec{e}_x \downarrow}\right)+\text {H.c.}\right]\\&+\left[\sum_{\vec{i}} \mathrm{i} t_{\mathrm{so}}\left({c}_{\vec{i} \uparrow}^{\dagger} {c}_{\vec{i}+\vec{e}_y \downarrow}-{c}_{\vec{i} \uparrow}^{\dagger} {c}_{\vec{i}-\vec{e}_y \downarrow}\right)+\text {H.c.}\right]\\&+\sum_{\vec{i}} \left((\Omega-\mathrm{i}\Omega){c}_{\vec{i \uparrow}}^{\dagger} {c}_{\vec{i}\downarrow}+\text {H.c.}\right)-\sum_{\vec{i}}\mathrm{i}\gamma{n}_{\vec{i} \downarrow}.
	\end{aligned}
	\label{eq:H}
\end{equation}
Here $c^\dagger_{\vec{i}\sigma}$ creates a particle at position $\vec{i}=(i_x,i_y)$ with spin $\sigma$, and $n_{\vec{i}\sigma}=c^\dagger_{\vec{i}\sigma}c_{\vec{i}\sigma}$. The unit vectors $\vec{e}_x=(1,0)$ and $\vec{e}_y=(0,1)$. The nearest-neighbour hopping rate is given by $t_s=\int d^2\mathbf{r}\phi_\uparrow^{0,0}(x,y)\frac{\mathbf{p}^2}{2m}+V(\mathbf{r})\phi_\uparrow^{0,0}(x-a,y)$.
The nearest-neighbour and on-site spin-flip rates are
$t_{so}=M_0\int d^2\mathbf{r}\phi_\uparrow^{0,0}(x,y)\sin(k_0x)\phi_\downarrow^{0,0}(x-a,y)$, $\Omega=\frac{\sqrt{2}}{2}M_r\int d^2\mathbf{r}\phi_\uparrow^{0,0}(x,y)e^{ik_0x}\phi_\downarrow^{0,0}(x,y)$.
Note that in the Hermitian limit with $\gamma=0$, our model differs from that in Ref.~\cite{shuai2d} by an additional on-site spin-flip term (characterized by $\Omega$), which is crucial for inducing the corner skin effect in the presence of loss.

Hamiltonian (\ref{eq:H}) has non-Hermitian skin effect in both the $x$ and $y$ directions under finite $\gamma$ and $\Omega$. In Fig.~\ref{fig:fig2}, we show typical eigenspectra and spatial distribution of eigenstates under different boundary conditions.
When the lattice has the periodic boundary condition along both spatial directions,
the eigenstates distribute uniformly across the 2D lattice [see Fig.~\ref{fig:fig2}(d)].
When the lattice has periodic boundary condition only along the $x$ direction, but has open boundary along the $y$ direction, the eigenstates are localized toward an open boundary[see Fig.~\ref{fig:fig2}(e)].
When both directions are under the open boundary condition,
the eigenstates are localized toward a corner [see Fig.~\ref{fig:fig2}(f)].
This is a signature of the non-Hermitian corner skin effect in 2D systems.
Further, from the eigenspectra on the complex plane, in-gap (line gap) topological edge states are clearly visible in Fig.~\ref{fig:fig2}(b)(c), when open boundaries are present. In the following, we characterize these topological edge states through the non-Bloch band theory.

\section{Topology under the non-Bloch band theory}

Under the NHSE, eigenstates deviate from extended Bloch waves.
Taking such deformation into account, we replace the phase factors $e^{\mathrm{i}k_x}$ and $e^{\mathrm{i}k_y}$ with the spatial mode factors $\beta_x(k_x)=|\beta_x(k_x)|e^{\mathrm{i}k_x}$ and $\beta_y(k_y)=|\beta_y(k_y)|e^{\mathrm{i}k_y}$~\cite{wangz1d,wangz2d}. Here the quasimomenta $k_x,k_y\in[0,2\pi)$, and the surface spanned by $\beta_x(k_x)$ and $\beta_y(k_y)$ is known as generalized Brillouin zone (GBZ) which can be calculated from the Schr\"{o}dinger’s equation as shown below.

We start from the non-Bloch Hamiltonian, formally derived by replacing $e^{\mathrm{i}k_x}$ ( $e^{\mathrm{i}k_y}$) with $\beta_x(k_x)$ ($\beta_y(k_y)$) in the dispersion relation of Hamiltonian (\ref{eq:H}) in the quasimomentum space (hence under the periodic boundary condition). The non-Bloch Hamiltonian is
\begin{align}		
		&H(\beta_x,\beta_y)=\nonumber\\
&\left[\Omega+ t_{\mathrm{so}}(\beta_{y}-\beta_{y}^{-1}) \right]\sigma_{x}+\left[\Omega+ t_{\mathrm{so}} (\beta_{x}-\beta_{x}^{-1})\right]\sigma_{y}\nonumber\\
&+\left[\Gamma_{z}+\mathrm{i}\frac{\gamma}{2}- t_{s}(\beta_{x}+\beta_{x}^{-1}+\beta_{y}+\beta_{y}^{-1})\right] \sigma_{z}-\mathrm{i}\frac{\gamma}{2}\mathbb{I},
	\label{eq:Hbeta}
\end{align}
where $\sigma_x$, $\sigma_y$ and $\sigma_z$ are the Pauli matrices, and $\mathbb{I}$ is the $2\times 2$ identity matrix. The corresponding Schr\"{o}dinger's equation is then $[H(\beta_x,\beta_y)-E_m]|\varphi^{R}_m(\beta_x,\beta_y)\rangle=0$, where
$E_j$ is the complex eigenenergy, $|\varphi^{R}_m(\beta_x,\beta_y)\rangle$ is the right eigenstate of $H(\beta_x,\beta_y)$ ($m=1,2$ being the band index). Sending the determinant of coefficients to zero, we have
\begin{align}		
&(E+\mathrm{i}\frac{\gamma}{2})^2-\left[\Omega+ t_{\mathrm{so}}(\beta_{y}-\beta_{y}^{-1}) \right]^2-\left[\Omega+ t_{\mathrm{so}} (\beta_{x}-\beta_{x}^{-1})\right]^2\nonumber\\
&-\left[\Gamma_{z}+\mathrm{i}\frac{\gamma}{2}- t_{s}(\beta_{x}+\beta_{x}^{-1}+\beta_{y}+\beta_{y}^{-1})\right]^2=0.
	\label{eq:betaeq}
\end{align}
Equation (\ref{eq:betaeq}) is a quartic equation for both $\beta_x$ and $\beta_y$.
Following the standard practice, we sort the four roots of Eq.~(\ref{eq:betaeq}) in ascending order $\{\beta_1,\beta_2,\beta_3,\beta_4\}$, and require $\beta_2=\beta_3$.
For each set $(k_x,k_y)$, we solve for the corresponding $|\beta_{x}(k_x)|$, $|\beta_y(k_y)|$ and $E$. While $E$ gives the eigenspectrum under the open boundary condition (along both directions), $|\beta_{x}(k_x)|$ and $|\beta_y(k_y)|$ determine the generalized Brillouin zone.

We are now in a position to calculate the non-Bloch Chern number, which can correctly predict the existence and number of the topological edge states.
The non-Bloch Chern number is defined as the surface integral of the Berry curvature over the generalized Brillouin zone
\begin{equation}	
	\begin{aligned}
		\mathcal{C}=-\frac{1}{2\pi}\int_{0}^{2\pi}\mathrm{d}k_x\int_{0}^{2\pi}\mathrm{dk}_y \mathbf{B}(k_x,k_y),
	\end{aligned}
	\label{eq:chernnum}
\end{equation}
where
\begin{equation}	
	\begin{aligned}
	 \mathbf{B}(k_x,k_y)=-\mathrm{Im}\frac{\langle\varphi_1^L|\nabla H|\varphi_2^R\rangle\times\langle\varphi_2^L|\nabla H|\varphi_1^R\rangle}{(E_1-E_2)^2}.
	\end{aligned}
	\label{eq:chernnum}
\end{equation}

Here the left eigenstate is defined as $H^\dagger(\beta_x,\beta_y)|\varphi_m^L(\beta_x,\beta_y)\rangle=E^*|\varphi_m^L(\beta_x,\beta_y)\rangle$. Note that for the Hermitian case $\gamma=0$, $\beta_x$ ($\beta_{y}$) is reduced to $e^{\mathrm{i}k_x}$ ($e^{\mathrm{i}k_y}$), and the non-Bloch Chern number to the conventional Bloch Chern number.

\begin{figure}[tbp]
	\includegraphics[width=9cm]{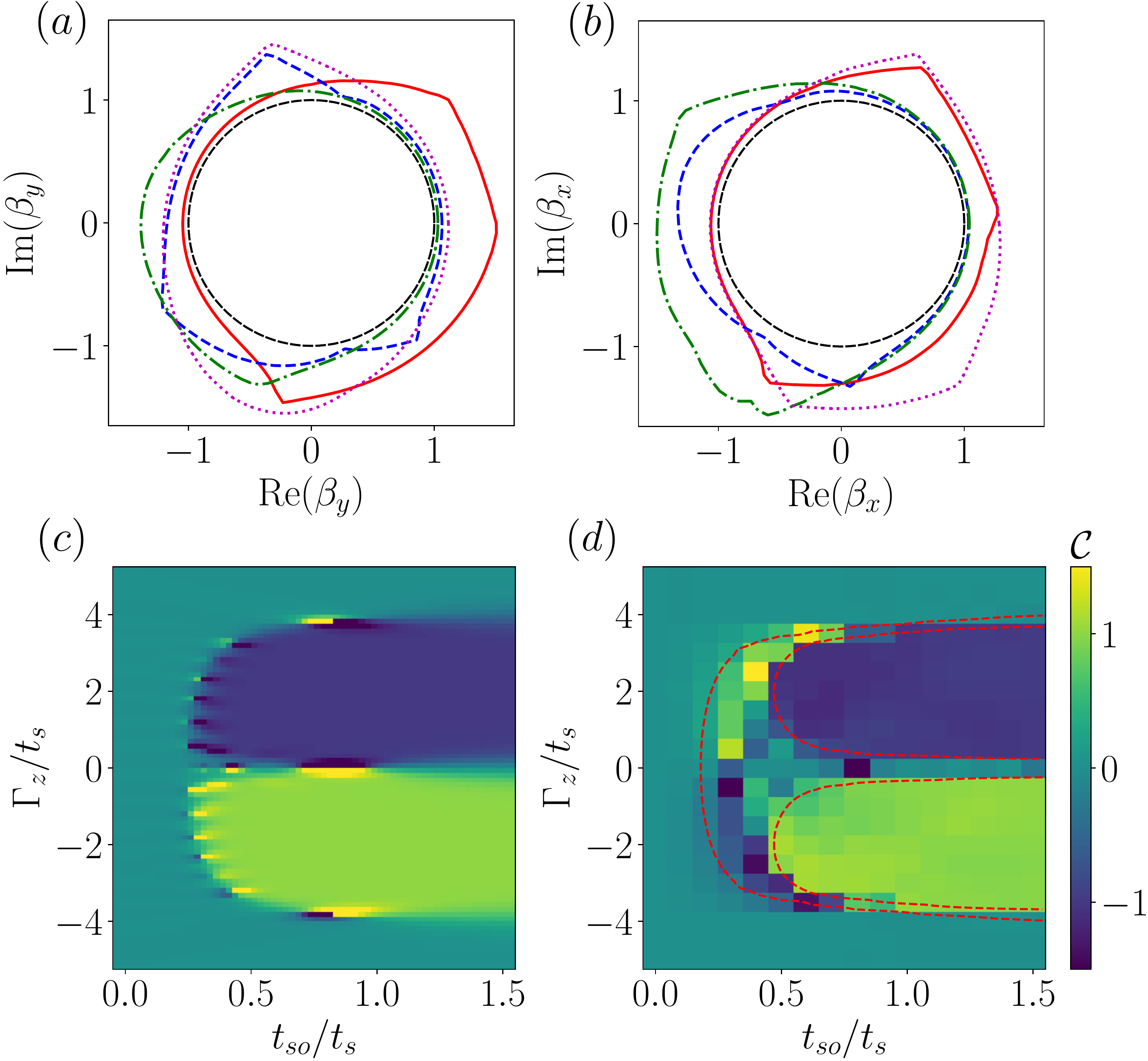}
	\caption{(a)(b) Cross sections of the generalized Brillouin zone on the complex plane, each with a fixed $k_x$ or $k_y$.
Red { solid}, blue { dashed}, green { dashdot}, and purple { dotted} loops indicate fixed $k_x=0,\pi/2,\pi,3\pi/2$ in (a) and $k_y=\pi/4,3\pi/4,5\pi/4,7\pi/4$ in (b). { Black densely dashed loop indicates the unit circle.} (c)(d) Topological phase diagrams of the lower band on the $t_{so}$--$\Gamma_{z}$ plane with (c) $\gamma=0$, and (d) $\gamma/t_s=2$. The red dashed lines in (d) denote the gapless region. In (a)(b), we take $\Gamma_{z}/t_s=0.5$ and  $t_{so}/t_s=1.5$, while for all figures, $\Omega/t_s=0.5$.}
	\label{fig:fig3}
\end{figure}

In Fig.~\ref{fig:fig3}(a)(b), we show the different cross sections of the generalized Brillouin zone by fixing $k_x$ or $k_y$. All of the loops have $|\beta_{x,y}|>1$, indicating that under the open boundary condition along both directions, all eignstates accumulate to the corner with large $j_x$ and $j_y$ indices.

In Fig.~\ref{fig:fig3}(c)(d), we plot the topological phase diagrams of the lower band for the Bloch and non-Bloch cases, respectively. In the Hermitian case [Fig.~\ref{fig:fig3}(c)], there are three distinct regions, respectively with $\mathcal{C}=0$ and $\mathcal{C}=\pm 1$. In the non-Hermitian case [Fig.~\ref{fig:fig3}(d)], the topological non-trivial phase regions with $\mathcal{C}=\pm 1$ become smaller, and a gapless region appear (bounded by red dashed lines) between the trivial and non-trivial phases. The non-Bloch Chern number cannot be defined in the gapless region.

\begin{figure*}[tbp]
	\includegraphics[width=18cm]{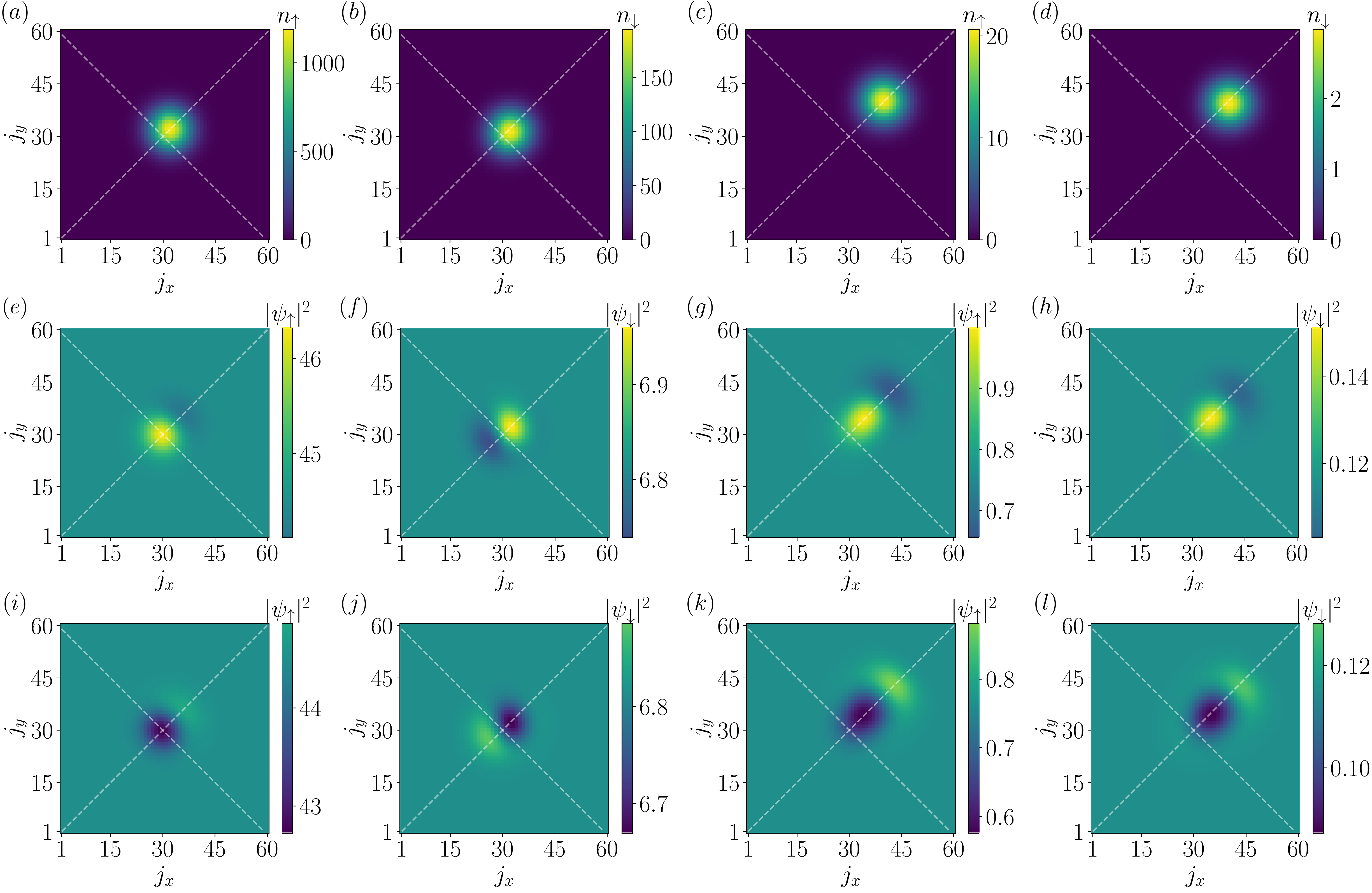}
	\caption{Directional dynamics of condensates in the Raman lattice. (a)(b)(e)(f)(i)(j) Contours of the density distribution for $\left|\uparrow\right\rangle$ and $\left|\downarrow\right\rangle$ states at $\tau t_s=2$. (c)(d)(g)(h)(k)(l) Contours of the density distribution for $\left|\uparrow\right\rangle$ and $\left|\downarrow\right\rangle$ states at $\tau t_s=10$. The initial state is the Gaussian wave packet state for (a)-(d), and the uniform state for (e)-(l). White dashed lines indicate the initial wave packet center for the lattice in (a)-(d), and for the impurity in (e)-(l). The interaction strengths are $u=-0.01$ for (e)-(h), and $u=0.01$ for (i)-(l), respectively. For all figures, lattice size $M=60$, $t_{so}/t_s=1$, $\Gamma_{z}/t_s=1$, $\Omega/t_s=2$, and $\gamma/t_s=2$.}
	\label{fig:fig4}
\end{figure*}

\section{Detecting skin effect through dynamics}

The NHSE generally leads to a directional flow in the system dynamics. This is visible by considering the dynamics of an initially quasi-localized condensate in the lattice space.

Specifically, we consider a condensate of spin-up atoms, loaded in a two-dimensional lattice with $M\times M$ sites. The initial condensate wave function is written as
\begin{equation}
|\psi(\tau=0)\rangle=\sqrt{N}\sum_{\vec{j}}\phi_{\vec{j}}c^\dagger_{\vec{j},\uparrow}|0\rangle,
\label{eq:psiG}
\end{equation}
where $\phi_{\vec{j}}^2=\exp(-|\vec{j}-\vec{j_0}|^2/w^2)/\pi w^2$, with $\vec{j_0}=(M/2,M/2)$ and $w=M/10$, and $\tau$ is the evolution time.
The initial atom number is $N=5\times 10^5$.
The time evolution is then given by $|\psi(\tau)\rangle=e^{-\mathrm{i}H\tau}|\psi(\tau=0)\rangle$.
Since we focus on the bulk dynamics, we consider only the short-time dynamics such that the condensate wavefunction has not yet evolved to the boundaries of the lattice.
We define the local density $n_{\vec{j},\sigma}=|\langle\vec{j},\sigma|\psi(\tau)\rangle|^2$, and then show the time-evolved density distribution in Fig.~\ref{fig:fig4}(a-d).
Apparently, both spin species exhibit a directional flow along the diagonal of the lattice, indicating the non-Hermitian corner skin effect.

In typical experiments, it is more natural to load atoms into the ground state of the Raman lattice, which under most parameters occurs at $k_x=k_y=0$. In the real space, this corresponds to an evenly distributed condensate wavefunction. Because of its homogeneity, as well as the lattice-translational symmetry, the directional bulk flow is not reflected in the time evolution of the density profiles. We therefore consider an alternative scheme where an impurity species is introduced, which interacts spin-selectively with atoms in the Raman lattice.
For convenience, we consider the impurity atoms to be trapped by a deep optical lattice potential with the same lattice constant as that of the Raman lattice, such that inter-sites hoppings are negligible (a finite hopping rate does not qualitatively change our discussion below).
The Hamiltonian concerning the impurity atoms then reads
\begin{equation}
	H_{\text{imp}}=t_{\text{imp}} \sum_{<\vec{i}, \vec{j}>}\hat{b}_{\vec{i }}^{\dagger} \hat{b}_{\vec{j }}+u\sum_{\vec{i}}\hat{b}_{\vec{i }}^{\dagger} \hat{b}_{\vec{i }}\hat{c}_{\vec{i} \uparrow}^{\dagger} \hat{c}_{\vec{i} \uparrow},
	\label{eq:Himp}
\end{equation}
where $b^\dagger_{\vec{i}}$ ($b_{\vec{i}}$) is the creation (annihilation) operator for the impurity atom at position $\vec{i}$, $t_{\text{imp}}$ is the nearest-neighbour hopping rate for the impurity, and $u$ is the spin-selective interaction strength.

We consider the case where a dissipative Bose-Einstein condensate is loaded into the Raman lattice.
This enables a mean-field treatment of the coupled dynamics. Specifically, we derive the equations of motion for the field operators $\{c_{\vec{j},\sigma},b_{\vec{j}}\}$ through
\begin{align}
i\hbar \frac{d}{dt} c_{\vec{j},\sigma}&=(H+H_{\text{imp}})c_{\vec{j},\sigma}-c_{\vec{j},\sigma}(H^\dag+H_{\text{imp}}),\\
i\hbar \frac{d}{dt} b_{\vec{j}}&=(H+H_{\text{imp}})b_{\vec{j}}-b_{\vec{j}}(H^\dag+H_{\text{imp}}).
\end{align}
We then take the mean-field approximation, defining the on-site condensate wavefunctions
$\psi_{\vec{j},\sigma}=\langle c_{\vec{j},\sigma}\rangle$ and $\psi_{\vec{j},\text{imp}}=\langle b_{\vec{j}}\rangle$. This allows us to numerically evolve the wave functions. For the initial state, we take
$\psi_{\vec{j},\sigma}=\sqrt{N_0} \xi_\sigma$ and $\psi_{\vec{j},\text{imp}}=\sqrt{N_{\text{imp}}}\sum_{\vec{j}}\phi_{\vec{j}}b^\dagger_{\vec{j}}|0\rangle$, where $\xi_\sigma$ is the ground state of Hamiltonian (\ref{eq:H}) under $\gamma=0$, and $\phi_{\vec{j}}$ has the same form as in (\ref{eq:psiG}).
The initial atom number of the condensate $N_0$ and the impurity species $N_{\text{imp}}$ are set to $5\times 10^5$ and $10^3$, respectively.

We show the results of the density evolution in Fig.~\ref{fig:fig4}(e-l). With an attractive interaction $u<0$, the impurity induces density peaks near $\vec{j_0}$ for both spin species.
The density peak in either spin species is accompanied by a dip in the other.
Because of these density excitations, the homogeneity of the initial state is broken
Driven by the directional flow, both density peaks propagate along the diagonal of the lattice, as shown in Fig.~\ref{fig:fig4}(g)(h).
The overall picture is similar under the repulsive interaction $u>0$, only the density peak (dip) under the attractive interaction is replaced with dip (peak), see Fig.~\ref{fig:fig4}(i-l).
Importantly, the propagation direction is independent of either spin species or the sign of the interaction, but related to the direction of the non-Hermitian corner skin effect.

\begin{figure}[tbp]
	\includegraphics[width=8.5cm]{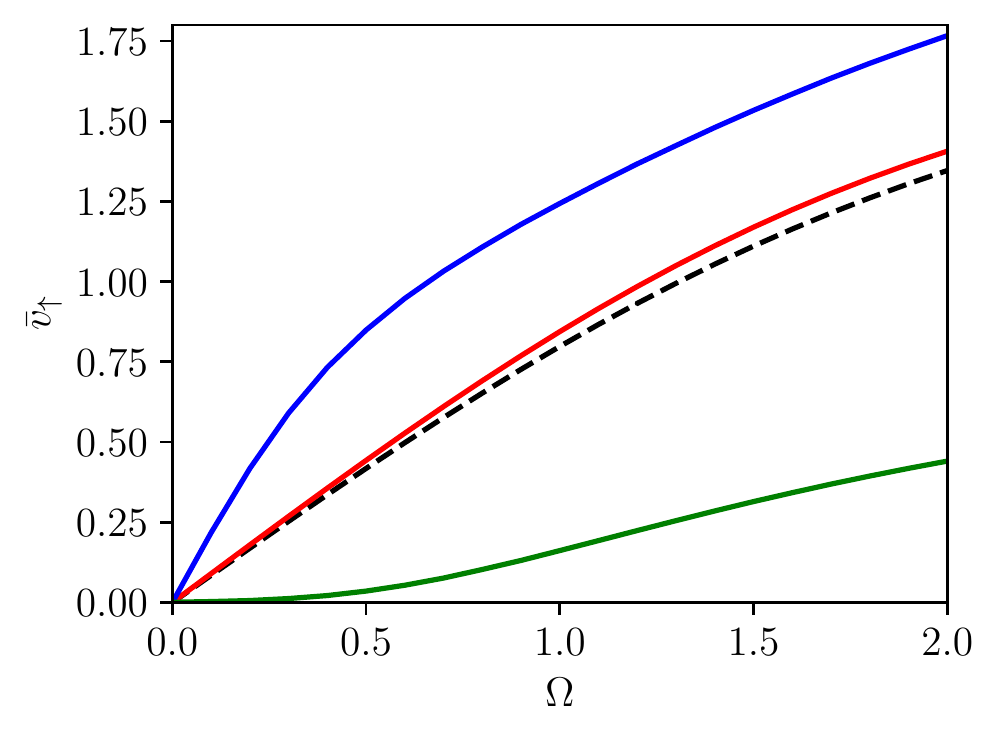}
	\caption{Mean peak velocity of spin-up atoms at $\tau t_s=10$. Red { (middling) solid line}: velocity corresponding to the cases of Fig.~\ref{fig:fig4}(a-c). Green { (lower) solid line}: velocity corresponding to the cases of Fig.~\ref{fig:fig4}(e-h). Blue { (upper) solid line}: velocity corresponding to the cases of Fig.~\ref{fig:fig4}(i-l). Black dashed line is the group velocity $v_g$ at $\vec{k}=\vec{0}$. }
	\label{fig:fig5}
\end{figure}

To investigate the effect of the interaction on the propagation of the wave packet, we define the mean peak velocity of spin-up atoms as
\begin{equation}
\bar{v}_{\uparrow}(\tau)=\frac{\sum_{\vec{j}}l(\vec{j})(|\psi_{\vec{j},\uparrow}|^2-n_{\rm{bg},\uparrow})\theta(|\psi_{\vec{j},\uparrow}|^2-n_{\rm{bg},\uparrow})}{\tau t_s\sum_{\vec{j}}(|\psi_{\vec{j},\uparrow}|^2-n_{\rm{bg},\uparrow})\theta(|\psi_{\vec{j},\uparrow}|^2-n_{\rm{bg},\uparrow})}.
	\label{eq:meanv}
\end{equation}
Here $l(\vec{j})=(j_x+j_y-M)/\sqrt{2}$ denotes the displacement in the diagonal direction. The background density $n_{\rm{bg},\uparrow}$ is defined as
$n_{\rm{bg},\uparrow}=|\psi_{(1,1),\uparrow}|^2$, as the excitations never reach the boundary during the evolution.

We plot the mean peak velocities for different cases in Fig.~\ref{fig:fig5}.
For the first scheme, (\ref{eq:meanv}) is reduced to $\bar{v}_{\uparrow}(\tau)=\sum_{\vec{j}}l(\vec{j})n_{\vec{j},\uparrow}/\tau t_s\sum_{\vec{j}}n_{\vec{j},\uparrow}$, and the mean peak velocity of the condensate (red) is very close to the group velocity $v_g$ at $\vec{k}=\vec{0}$ (black dashed)~\cite{nhsedy3}. Here the group velocity has the form $v_g=\left|\frac{d \mathrm{Re}(E^-)}{d\vec{k}}\right|$, where $E^-$ is the lower-band dispersion of the Raman lattice.
However, with interactions, the mean peak velocity shows a large deviation from the group velocity. The attractive (repulsive) interaction makes the evolution of the peak slower (faster), which is intuitive under the mean-field approximation.

\section{Conclusion}
We have shown that a two-dimensional dissipative Raman lattice features non-Hermitian corner skin effect and band topology, and is amenable to current cold-atom experiments. Two dynamic detection schemes for the NHSE are then proposed, based on a two-component Bose-Einstein condensate in the Raman lattice. In particular, we show that, by introducing an impurity species that interacts spin-selectively with a lattice condensate in the ground state, the directional flow, characteristic of the NHSE, is imposed upon the interaction-induced density excitations in the condensate. The interplay of interaction and NHSE can then be studied through the dynamics of the density excitations.
For future studies, it would be desirable to engineer more exotic higher-order NHSE based on the dissipative Raman lattice. The impurity-induced density excitations can also be studied in the context of polarons in a dissipative condensate, where the interplay of interaction and NHSE can be understood in the context of quasiparticle excitations.

 \begin{acknowledgments}
This work has been supported by the Natural Science Foundation of China (Grant Nos. 11974331) and the National Key R\&D Program (Grant No. 2017YFA0304100).
\end{acknowledgments}


\begin{thebibliography}{99}
    \bibitem{xjliu1d} X.-J. Liu, Z.-X. Liu and M. Cheng, Phys. Rev. Lett. {\bf 110}, 076401 (2013).
	\bibitem{xjliu2d} X.-J. Liu, K.T. Law, and T.K.Ng, Phys. Rev. Lett. {\bf 112}, 086401 (2014).
	\bibitem{shuai2d} Z. Wu, L. Zhang, W. Sun, X.-T. Xu, B.-Z.Wang, S.-C. Ji,
	Y. Deng, S. Chen, X.-J. Liu, and J.-W. Pan, Science {\bf 354}, 83 (2016).
	\bibitem{topoins2} B. Song, L. Zhang, C. He, T. F. J. Poon, E. Hajiyev, S. Zhang, X.-J. Liu, and G.-B. Jo, Sci. Adv. {\bf 4}, eaao4748 (2018).
\bibitem{ti1} M. Z. Hasan and C. L. Kane, Rev. Mod. Phys. {\bf 82}, 3045 (2010).

	\bibitem{shuai3d} Z.-Y. Wang,  X.-C. Cheng, B.-Z. Wang, J.-Y. Zhang, Y.-H. Lu,
	C.-R. Yi, S. Niu, Y. Deng, X.-J. Liu, S. Chen, and J.-W. Pan, Science {\bf 372}, 271-276 (2021).
	\bibitem{weyl2} H. Weng, C. Fang, Z. Fang, B. A. Bernevig, and X. Dai, Phys. Rev. X
	{\bf 5}, 011029 (2015).
\bibitem{dynRa1} C.-R. Yi, L. Zhang, L. Zhang, R.-H. Jiao, X.-C. Cheng, Z.-Y. Wang, X.-T. Xu, W. Sun, X.-J. Liu, S. Chen, and J.-W. Pan, Phys. Rev. Lett. {\bf 123}, 190603 (2019).
\bibitem{dynRa2} L. Zhang, L. Zhang, and X.-J. Liu, Phys. Rev. Lett. {\bf 125}, 183001 (2020).
\bibitem{dynRa3} D.-H. Cai and W. Yi, Phys. Rev. A {\bf 105}, 042812 (2022).
\bibitem{dissRa} L. Zhou, H. Li, W. Yi, and X. Cui, Commun. Phys. {\bf 5}, 252 (2022).

	\bibitem{zoller08} S. Diehl, A. Micheli, A. Kantian, B. Kraus, H. P. B\"{u}chler, and P. Zoller
	Nat. Phys. {\bf 4}, 878–883 (2008).
	\bibitem{blatt13} P. Schindler, M. Müller, D. Nigg, J. T. Barreiro, E. A. Martinez, M. Hennrich, T. Monz, S. Diehl, P. Zoller, and R. Blatt
	Nat. Phy. {\bf 9}, 361–367 (2013).
	\bibitem{zoller12}M. M\"{u}ller, S. Diehl, G. Pupillo, and P. Zoller, Adv. At.
	Mol. Opt. Phys. {\bf 61}, 1 (2012).
	\bibitem{disspt1}E. Fiorelli, I. Lesanovsky, and M. M\"{u}ller, New. J. Phys. {\bf 24}, 033012 (2022).
	\bibitem{disspt2}Y. Zhang and T. Barthel, Phys. Rev. Lett. {\bf 129}, 120401 (2022).
	\bibitem{disspt3}J. M. Fink, A. Dombi, A. Vukics, A. Wallraff, and P. Domokos, Phys. Rev. X {\bf 7}, 011012 (2017).
	\bibitem{disspt4}S. R. K. Rodriguez, W. Casteels, F. Storme, N. Carlon Zambon, I. Sagnes, L. Le Gratiet,
	E. Galopin, A. Lema\^{i}tre, A. Amo, C. Ciuti, and J. Bloch, Phys. Rev. Lett. {\bf 118}, 247402 (2017).
	\bibitem{disspt5} T. Fink, A. Schade, S. Höfling, C. Schneider, and A. Imamoglu, Nat. Phys. {\bf 14}(4), 365 (2018).
	\bibitem{Non1} N. Moiseyev, Non-Hermitian quantum mechanics, {\it Cambridge University Press} (2011).
	\bibitem{Uedareview} Y. Ashida, Z. Gong, and M. Ueda, Adv. Phys. \textbf{69}, 3 (2020).
	\bibitem{molmer}J. Dalibard, Y. Castin, and K. M\o lmer, Phys. Rev. Lett. {\bf68}, 580 (1992).
	\bibitem{michael}H. J. Carmichael, Phys. Rev. Lett. {\bf70}, 2273 (1993).
	\bibitem{weimer} H. Weimer, A. Kshetrimayum, and R. Or{\'u}s, Rev. Mod. Phys. {\bf93}, 015008 (2021).
	\bibitem{PT1}
	C. M. Bender and S. Boettcher, Phys. Rev. Lett. \textbf{80}, 5243 (1998).
	\bibitem{photonics2} R. El-Ganainy, K. Makris, M. Khajavikhan, Z. H. Musslimani, S. Rotter, and D. N. Christodoulides, Nat. Phys. \textbf{14}, 11-19 (2018).
	\bibitem{nhtopot1}Z. Gong, Y. Ashida, K. Kawabata, K. Takasan, S. Higashikawa, and M. Ueda, Phys. Rev. X {\bf 8}, 031079 (2018).
	\bibitem{nhtopot2}K. Kawabata, K. Shiozaki, M. Ueda, and M. Sato, Phys. Rev. X {\bf 9}, 041015 (2019).
	\bibitem{nhtopot3}D. Leykam, K. Y. Bliokh, C. Huang, Y. D. Chong, and F. Nori.	Phys. Rev. Lett. {\bf 118}, 040401 (2017).
	\bibitem{nhtopoe1}J. M. Zeuner, M. C. Rechtsman, Y. Plotnik, Y. Lumer, S. Nolte, M. S. Rudner, M. Segev, and A. Szameit, Phys. Rev. Lett. {\bf 115}, 040402 (2015).
	\bibitem{nhtopoe15} X. Zhan, L. Xiao, Z. Bian, K. Wang, X. Qiu, B. C. Sanders, W. Yi, and P. Xue, Phys. Rev. Lett. {\bf 119}, 130501 (2017).
	\bibitem{nhtopoe16} L. Xiao, X. Zhan, Z. Bian, K. K. Wang, X. Zhang, X. P. Wang, J. Li, K. Mochizuki, D. Kim, N. kawakami, W. Yi, H. Obuse, B. C. Sanders, and P. Xue, Nat. Phys. {\bf 13}, 1117 (2017).
\bibitem{luole} J. Li, A. K. Harter, J. Liu, L. de Melo, Y. N. Joglekar, L. Luo, Nat. Commun. {\bf 10}, 855 (2019).
\bibitem{yanbo1} W. Gou, T. Chen, D. Xie, T. Xiao, T.-S. Deng, B. Gadway, W. Yi, and B. Yan, Phys. Rev. Lett. {\bf 124}, 070402 (2020).
\bibitem{yanbo2} Q. Liang, D. Xie, Z. Dong, H. Li, H. Li, B. Gadway, W. Yi, and B. Yan, Phys. Rev. Lett. {\bf 129}, 070401 (2022).
\bibitem{joep}
Z.-J. Ren, D. Liu, E.-T. Zhao, C.-D. He, K. K. Pak, J. Li, and G.-B. Jo, Nat. Phys. \textbf{18}, 385 (2022).


	\bibitem{wangz1d}S. Yao and Z. Wang, Phys. Rev. Lett. {\bf 121},
	086803 (2018).
	\bibitem{wangz2d}S. Yao, F. Song, and Z. Wang, Phys. Rev. Lett. {\bf 121}, 136802 (2018).
	\bibitem{murakami}K. Yokomizo and S. Murakami, Phys. Rev. Lett. {\bf 123}, 066404 (2019).
	\bibitem{nhse1}C. H. Lee and R. Thomale, Phys. Rev. B \textbf{99}, 201103(R) (2019).
	\bibitem{nhse2}A. McDonald, T. Pereg-Barnea, and A. A. Clerk, Phys. Rev. X \textbf{8}, 041031 (2018).
	\bibitem{nhse3}K. Zhang, Z. Yang, and C. Fang, Phys. Rev. Lett. \textbf{125}, 126402 (2020).
	\bibitem{nhse4}N. Okuma, K. Kawabata, K. Shiozaki, and M. Sato, Phys. Rev. Lett. \textbf{124}, 086801 (2020).
	\bibitem{nhse5}T.-S. Deng and W. Yi, Phys. Rev. B \textbf{100}, 035102 (2019).
	\bibitem{nhse6}Z. Yang, K. Zhang, C. Fang, and J. Hu, Phys. Rev. Lett. \textbf{125}, 226402 (2020).
	\bibitem{nhsedy1}S. Longhi, Phys. Rev. Research \textbf{1}, 023013 (2019).
	\bibitem{nhsedy2}T. Li, J.-Z. Sun, Y.-S. Zhang, and W. Yi, Phys. Rev.
	Research \textbf{3}, 023022 (2021).
	\bibitem{nhsedy3}S. Longhi, Phys. Rev. B \textbf{102}, 201103(R) (2020).
	\bibitem{nhtopoe2}L. Xiao, T. Deng, K. Wang, G. Zhu, Z. Wang, W. Yi, and P. Xue, Nat. Phy. {\bf 16}, 761–766 (2020).
	\bibitem{scienceskin} S. Weidemann, M. Kremer, T. Helbig, T. Hofmann, A. Stegmaier, M. Greiter, R. Thomale, and A. Szameit, Science {\bf 368}, 311 (2020).
	\bibitem{2dnhseacoustics}X. Zhang, Y. Tian, J.-H. Jiang, M.-H. Lu, and Y.-F. Chen, Nat. Commun. {\bf 12}, 5377 (2021).
	\bibitem{2dnhsecircuits}D. Zou, T. Chen, W. He, J. Bao, C. H. Lee, H. Sun, and X. Zhang, Nat. Commun. {\bf 12}, 7201 (2021).
\bibitem{qw1} Q. Lin, T. Li, L. Xiao, K. Wang, W. Yi, and P. Xue, Nat. Commun. {\bf 13}, 3229 (2022).
\bibitem{qw2} Q. Lin, T. Li, L. Xiao, K. Wang, W. Yi, and P. Xue, Phys. Rev. Lett. {\bf 129}, 113601 (2022).
\bibitem{cuisoc} L. Zhou, W. Yi, and X. Cui, Phys. Rev. A {\bf 102}, 043310 (2020).
\end{thebibliography}
\end{document}